# MD Simulations of Compression of Nanoscale Iron Pillars

Con J. Healy and Graeme J. Ackland
School of Physics and Astronomy, The University of Edinburgh,
James Clerk Maxwell Building, Mayfield Road, Edinburgh EH9 3JZ, UK

## ABSTRACT

It is now possible to create perfect crystal nanowires of many metals. The deformation of such objects requires a good understanding of the processes involved in plasticity at the nanoscale. Isotropic compression of such nanometre scale micropillars is a good model system to understand the plasticity. Here we investigate these phenomena using Molecular Dynamics (MD) simulations of nanometre scale single crystal BCC iron pillars in compression.

  We find that pillars with large length to width ratio may buckle under high strain rates. The type of buckling behaviour depends sensitively on the boundary conditions used: periodic boundary conditions allow for rotation at top and bottom of the pillar, and result in an S shaped buckle, by contrast fixed boundaries enforce a C shape. Pillars with a length to width ratio closer to that used in experimental micropillar compression studies show deformation behaviour dominated by slip, in agreement with the experiments. For micropillars oriented along <100>, slip occurs on <110> planes and localized slip bands are formed. Pillars of this size experience higher stresses than bulk materials before yielding takes place. One might expect that this may be in part due to the lack of nucleation sites needed to induce slip. However, further simulations with possible dislocation sources: a shorter iron pillar containing a spherical grain boundary, and a similar pillar containing jagged edges did not show a decreased yield strength.

## INTRODUCTION

Despite advances made over the course of the last century on microscopic behavior of solids, the microstructure of solids remains difficult to represent in models of macroscopic plasticity. New experimental techniques are now becoming available to analyze deformation at smaller length scales. The micropillar compression test is an experimental technique which has been developed to analyze plastic deformation at the micron scale [1]. In this technique, pillars of single crystal metal are created and then compressed using an indenter. These pillars then usually deform by slip. Slip is a deformation mechanism characterized by closely packed planes in a crystalline material slide across each other as dislocations pass through the material.

  Results from pillar compression tests of FCC micropillars show significant size scale dependence on the strength of the micropillar, with smaller micropillars being stronger [1]. Since a micropillar is likely to present less of an obstacle to the movement of dislocations, the implication is that the strength comes from the difficulty in nucleation of sufficient dislocations, combined with their easy destruction at the surface. Molecular Dynamics simulations of FCC nanopillars by Li & Yang [2] show further size scale dependence on strength of pillars on the nanometre scale. This affirms size scale effects recorded elsewhere in studies of single crystal nanowires [3].

  In this paper, we outline MD simulations of nanoscale BCC iron pillars in compression.

Technical modeling issues such as the effect of periodic boundary conditions are shown to be important. The effect of defects such as grain boundaries is also investigated.

**SIMULATIONS**

Simulations were carried out using the MOLDY molecular dynamics package [4]. The pairwise potential function used was the EAM-type potential for iron developed by Hepburn and Ackland [5]. In all simulations, nanoscale pillars were constructed from iron atoms in a BCC crystalline structure.

**Buckling of long nanoscale pillars**

A simulation was carried out which consisted of a pillar of iron atoms with a square cross-section. The pillar consisted of 50520 atoms and had an initial length-to-width ratio of 8:1. Consequently, the pillar was approximately 4.3nm in width and 34.3nm in length. Periodic boundary conditions were implemented along the z-axis which ran parallel to the long side of the pillar. The pillar was oriented so that the z-axis ran parallel to the long sides of the pillar. The long faces of the pillar were free surfaces. This was achieved by implementing periodic boundary conditions in the x and y directions but constructing the pillar such that its base covered only a quarter of the area of the x-y plane of the repeated cell. The pillar was oriented such that the z direction pointed in the <001> direction of the crystal. The pillar was compressed by rescaling the z coordinates of the atomic positions by 0.05% at intervals of 200fs. This led to a strain rate of $2.5 \times 10^9 s^{-1}$ in the <001> direction. This strain rate is much higher than strain rates observed experimentally [1]. However, this is inevitable due to the small timescales over which MD simulations are typically carried out. This temperature was set at zero Kelvin initially with no thermostat used: the pillar warms up adiabatically as strain energy is converted to heat. A Predictor-Corrector algorithm was used to integrate the atomic equations of motion with a 1fs time-step.

The disordered nature of the compressed pillar made it difficult to identify incidences of twinning. An analysis of nearest neighbor changes showed that slip had not taken place during pillar compression. A large amount of plastic deformation was observed during compression of these pillars but this appeared to occur by movement. After a strain of just over 20%, the pillar buckled. An image of the buckled pillar can be seen in Figure 1.

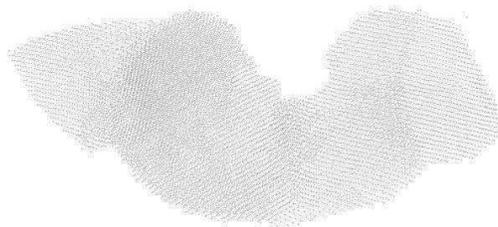

*Figure 1:  Buckled nanopillar forming an "S" shape*

Figure 2 shows a graph of load force on the faces of the cell against the engineering strain. The load force is the force on the top and bottom surfaces of the pillar as it is compressed.

This is related to the stress in pillar as the stress is pressure on the top and bottom surfaces of the pillar. However, these two quantities are not linearly proportional throughout the simulation as the area on the top and bottom surfaces of the pillar change over the course of the simulation. It is clear from this graph that the pillar experiences stresses far higher than the yield stress of bulk iron. This may be attributed to the lack of grain boundaries and dislocations in the pillar and the high strain rates experienced during the simulation. Simulations of FCC nanopillars and nanowires in compression show stresses in the order of GPa and tens of GPa. However, strains of 20% without slip or twinning is unusual or unobserved in studies of FCC crystals in compression [2, 6, 7].

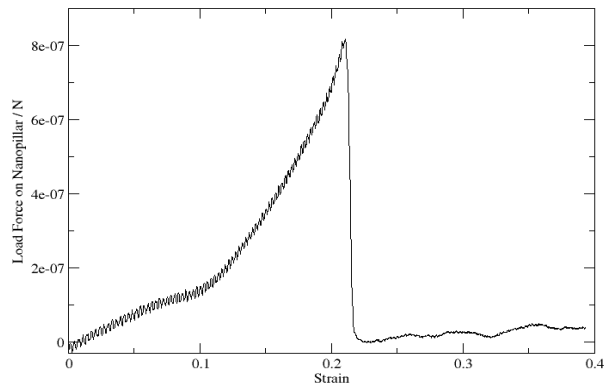

*Figure 2: Load force vs. strain for buckled pillar*

The type of buckling behavior depends sensitively on the boundary conditions used. The pillar in Figure 1 bends at two points to form an "S" shape. This may be attributed to a conservation of momentum effect. If the pillar were to bend at only one point and form a "C" shape, there would be a net displacement in the pillar, which can have occurred only if there was a net change in momentum at some point. Furthermore, the top and bottom surfaces of the pillar are rotated, but parallel to one another as required by the periodic boundary conditions.

An alternate method of applying strain is to have two layers of atoms on the top and bottom of the pillar constrained to move as rigid bodies. The effective forces keeping these atoms in motion provide the stress. The pillar in this simulation behaved in a similar way. Deformation occurred again, not through twinning or slip, but through the movement of individual atoms. Again, the pillar eventually buckled. Figure 3 shows an image of the buckled pillar from this simulation. The pillar forms a "C" shape with the pillar bending at just one point.

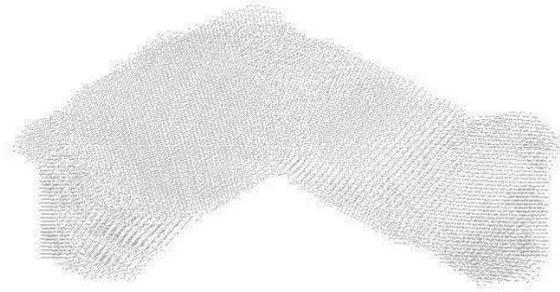

*Figure 3: Buckled pillar forming a "C" shape.*

**Slip bands in shorter wider pillars**

Shorter, wider pillars showed different deformation mechanisms when compressed. For this simulation a pillar 15.4nm in length and 5.7nm in width was constructed. The top and bottom faces of the pillar were connected to square plates of atoms approximately 3.08nm in thickness and 11.4nm in width. The crystal orientation of the pillar was the same as that of the long narrow pillars. The simulation contained 59055 atoms in total. The time-step used was 1fs and the Verlet integration algorithm[8] was used to integrate the equations of motion. The length of this pillar was rescaled by at intervals of 2000fs. This led to a strain rate of $2.5 \times 10^8 s^{-1}$. Figure 4 shows an image of this pillar with an applied strain of 18%. Figure 5 shows the same pillar at the same strain except only atoms for which nearest neighbors changed significantly are shown. The pillar has deformed via slip on <110> planes. These slip events appear to have occurred in a localized band.

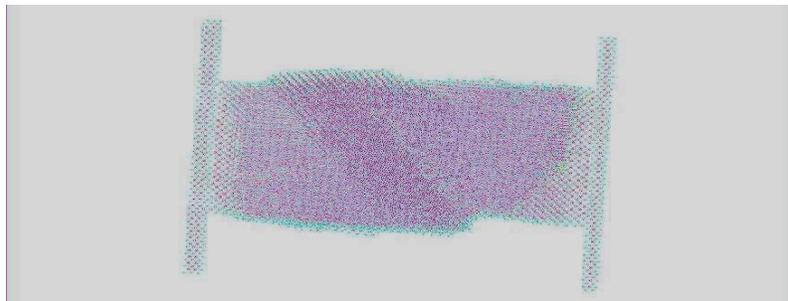

*Figure 4: Nanopillar with applied strain of 18%. A slip band is visible in the center of the pillar.*

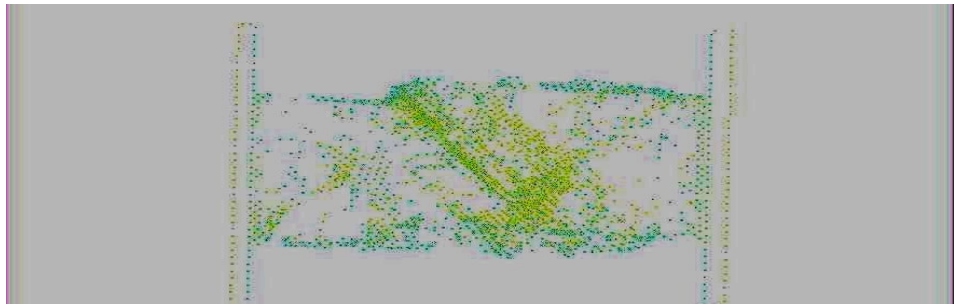

*Figure 5: Nanopillar with applied strain of 18%. Only atoms for which nearest neighbors changed significantly are shown*

Similar simulations were carried out with the addition of defects to this pillar. The above simulation was repeated with the addition of a spherical grain boundary to pillar. This grain boundary was created by rotating a sphere of atoms in the center of the pillar through an arbitrary angle. Some atoms at the boundary surface were then removed to prevent any atoms from being unrealistically close together. The diameter of the spherical grain boundary was approximately 2.9nm. The simulation was also repeated after adding a jagged edge to the pillar. Both of these nanopillars also deformed through the formation of a slip band. Figure six shows a graph of load force vs. strain. This graph reveals a stress-strain behavior in the pillars characterized gradual increases in the stress followed by a sudden lowering of the stress values. These sudden drops in the stress can be attributed to slip events. One can see from figure 6 that the addition of a jagged edge to the pillar made little difference to the yield stress while the addition of the grain boundary increased the yield stress.

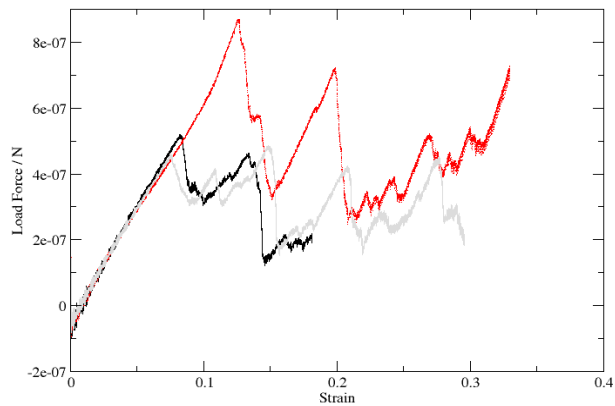

*Figure 6: Load force vs. strain for different nanopillars. The values for the initially defect free pillar are shown in black while the values for the pillar with a jagged edge are shown in light grey and the values for the nanopillar with a spherical grain boundary are represented by the dotted red line.*

## DISCUSSION AND CONCLUSIONS

Molecular dynamics of nanoscale iron pillars in compression in the <001> direction show that long pillars appeared to buckle before slip or twinning occurred. The shape of these pillars after buckling was influenced by periodic boundary conditions. This is unlike reported experimental deformation of micropillars.
    Simulations the shorter pillars recreated some aspects of micropillar compression experiments: slip was observed and appeared in bands. While it was not possible to identify the character of individual dislocations, the banding suggests that the come from a related source.
    Unlike micropillar compression experiments, the stress did not remain constant after the initial yield. This can be attributed to stochastic effects from the small size of the pillar: each dip

in the stress-strain graph represents a single yielding event.

One might expect that adding defects would facilitate creating a nucleation site for dislocations or a dislocation source and therefore decrease the yield stress. However, the defects added to the shorter pillar did not have this effect. In the case of the pillar with the grain boundary the initial yield stress actually increased. This may have been due to the spherical grain impeding slip motion, and failing to act as a dislocation source.

The yield stresses in these pillars were of the order of several GPa, much higher than those typically seen in micropillar compression experiments. However, micropillar compression studies show increasing values for yield stress for smaller pillars which makes the yield stresses seen in these simulations seem more plausible.

## ACKNOWLEDGMENTS

The research leading to these results has received funding from the European Union Seventh Framework Programme (FP7/2007-2013) under grant agreement No. PITN-GA-2008-211536, project MaMiNa.